\documentclass[runningheads]{llncs}
\usepackage{graphicx}
\usepackage[resetlabels,labeled]{multibib}
\newcites{R}{Selected Papers}

\begin{document}
\title{Characteristics of an Online Controlled Experiment: Preliminary Results of a Literature Review}
\titlerunning{Characteristics of an Online Controlled Experiment}
\author{Florian Auer\inst{1} \and
Michael Felderer\inst{1}}
\authorrunning{F. Auer et al.}
\institute{University of Innsbruck, 6020 Innsbruck, Austria
\email{\{florian.auer,michael.felderer\}@uibk.ac.at}}
\maketitle              
%
\begin{abstract}
In this paper the preliminary results of a literature review on characteristics used to define continuous experiments are presented. In total 14 papers were selected. The results were synthesized into a model that gives an overview of all characteristics.

\end{abstract}
%
%
%

\section{Motivation}
The body of knowledge on continuous experimentation increases steadily \cite{auer2018current}. However, with increasing number of publications on experimentation, it becomes more difficult for practitioners and researchers to keep track of all advances in the field. One fundamental question that requires an holistic overview on the body of knowledge is about the characteristics of an experiment definition. What characteristics are known of experiments that could be used to define experiments? The answer of this question is of importance for practitioners to define or improve their process of experimentation. Therefore, the following research question will be discussed in this paper:\\

\noindent\textbf{RQ: What characteristics of online controlled experiments are described in the body of knowledge of continuous experimentation
to define experiments?}

\section{Research Method}
The body of knowledge on continuous experimentation was reviewed to find out which characteristics of experiment definitions are described in literature. Given that it is not possible to came up with the characteristics to search for prior the analysis, a search term based literature lookup was not possible. 

Therefore, the papers selected by two published literature reviews in the field of continuous experimentation \cite{auer2018current,ros2018continuous22} were considered as starting set. They represent the body of knowledge in this field and thus are a good set of papers to start from. The two reviews selected 82 \cite{auer2018current} and 62 \cite{ros2018continuous22} papers. As both studies are not primarily focused on the definition of experiments, the papers not relevant for this study were removed by applying the following inclusion criteria on the title, abstract, conclusions and in doubt by skimming through the paper:

\begin{itemize}
    \item Explicit description of experiment characteristics
    \item Guidelines on defining experiments, how to design experiments
    \item Checklist prior running an experiment
    \item Experiment review process or guidelines
    \item Characterization of experimentation
\end{itemize}

The inspection of the papers resulted into the removal of most papers. As a result nine papers remained in the set of relevant papers. Based on these papers a forward snowballing was applied, with the intend to find more relevant papers. To find all publications referencing a publication, the search engine Google Scholar\footnote{\url{https://scholar.google.com}} was used. One additional criteria was applied in the forward snowballing, namely that only publications published after 2017 were considered. This was done because the starting set composed of the two studies, were collected in 2018. The forward snowballing revealed five additional relevant papers. Thus, in total 14 papers were selected. The papers are listed in the bibliography (see \emph{Selected Papers}).

After the selection of the relevant papers, the publications were analysed in detail and relevant characteristics of experiment specification were extracted.

\section{Results}

\begin{figure}[h]
    \centering
    \includegraphics[width=\textwidth,page=1,trim=4cm 0cm 4cm 0cm,clip]{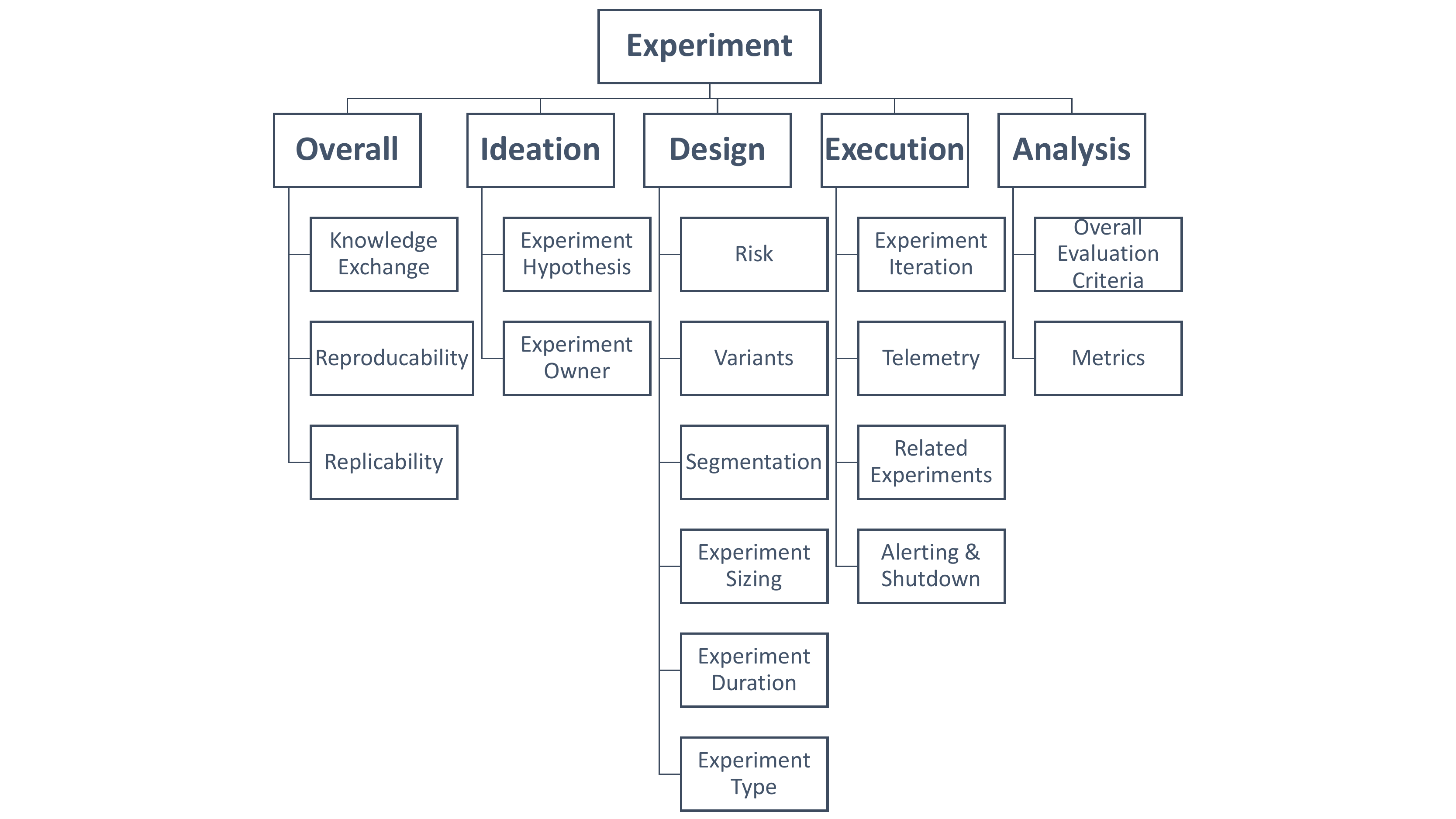}
    \caption{Extracted experiment definition characteristics from literature.}
    \label{fig:literature-characteristics}
\end{figure}

In the following the results of the literature review based on the 14 selected papers is presented. The findings are synthesized into a model (see Figure \ref{fig:literature-characteristics}), in which they are grouped according to the phases of the experiment lifecycle as proposed by Fabijan et al. \cite{fabijan2018online}. In addition, some characteristics were found to be of importance for every phase of the experiment (e.g. documentation) and thus added in each phase.

\section{Conclusions}

The primarily results of the literature review gives an overview of all in literature described characteristics of continuous experiments. The model (see Figure \ref{fig:literature-characteristics}) is of value for practitioners to give them an assembled overview on all characteristics that could be considered for the description of experiments. Researchers, on the other hand, can use the model to get an overview of the areas of experimentation and their characteristics, for example. 

In future work, the model will be extended with valuable information from practitioners about the characteristics importance for reliable experiments. This will further guide practitioners in the selection of characteristics for their concrete use cases.

\bibliographystyle{splncs04}
\bibliography{ref_primary}

\nociteR{*}
\bibliographystyleR{splncs04}
\bibliographyR{ref_review}
\end{document}